\documentclass[prd,reprint,showpacs,showkeys]{revtex4}
\usepackage{amsfonts}
\usepackage{amssymb}
\usepackage{amsmath}
\usepackage{graphicx}
\usepackage[font={footnotesize,it}]{caption}

\usepackage{graphicx}
\usepackage[colorlinks=true,
            linkcolor=blue,
            urlcolor=blue,
            citecolor=red]{hyperref}

\begin{document}

\title{Interpolating the Schwarzschild and de-Sitter metrics}
\author{M. Halilsoy}
\email{mustafa.halilsoy@emu.edu.tr}
\author{S. Habib Mazharimousavi}
\email{habib.mazhari@emu.edu.tr}
\affiliation{Physics Department, Arts and Sciences Faculty, Eastern Mediterranean
University, Famagusta, North Cyprus via Mersin 10, Turkey.}

\date{\today }

\begin{abstract}
The binary potential technique of interpolation (by M. Riesz, Acta Math. 81,
1 (1949)) is applied to some well-known metrics of general relativity. These
include Schwarzschild, de Sitter and 2+1-dimensional BTZ spacetimes. In
particular, the Schwarzschild-de Sitter solution is analyzed in some detail
with a finite range parameter. Reasoning by the high level of non-linearity
and absence of a superposition law necessitates search for alternative
approaches. We propose the method of interpolation between different
spacetimes as one such possibility paving the way toward controlling the
two-metric system by a common parameter.
\end{abstract}

\maketitle
\section{Introduction}

The renowned Majumdar-Papapetrou solution \cite{1,2,3,4,5,6} describes a
multi-black hole solution in which the electromagnetic and gravitational
forces balance each other. The black holes are not located arbitrarily but
rather they lie on an axis along which the attractive / repulsive force act.
Many-body problems, even the two-body, have always been tough in physics and
general relativity doesn't provide an exception in this regard. An important
lesson we learned from quantum theory is; particles that were connected /
interacted in the past by some mechanism remain ever connected in the future
albeit through a spooky action at a distance. This is known as the
Einstein-Podolsky-Rosen (EPR) paradox \cite{7}, keeping things entangled
through spacetime. Is there an analogue picture in classical relativity
where particles are to be replaced by spacetimes? In other words, can we
connect / couple two different spacetimes by a continuous parameter such
that at one end it yields the first while at the other end the second
spacetime? Although mathematically this is the process of interpolating two
different solutions it may be considered as a classical analogue of
entanglement for the two spacetimes. It was argued \cite{R1} (and references
cited therein) that the two distant particles may be connected through a
spacetime wormhole. A wormhole is a solution to the equations of classical
general relativity obeying the chronology protection whereas a spooky
connection violates causality. How to reconcile a classical wormhole with a
spooky quantum action? Further, unless one resorts to the quantum
fluctuations at the Planck scale the wormholes constructed \ from
Schwarzschild-related black holes are not traversable. Our motivation
originated from the fact that within classical theory we must find a way to
imitate the non-local effects. We aim to do this by introducing a family of
metrics governed by one (or more) control / interpolation parameter. The
method may be considered as an alternative to the one of wormhole
connection. The examples that we discuss here are the Schwarzschild (S) \cite%
{8}, Ba\~{n}ados-Teitelboim-Zanelli (BTZ) \cite{9}, and de Sitter \cite%
{10,11} spacetimes. We note that there have been previous examples that used
to interpolate different spacetimes \cite{12}. Let us draw a rough analogy.
In a classical two-bit system (say A and B) we are confronted with the
simple rule: either A or B with no gray zone. In a quantum system of qubits
on the other hand we have infinite possibilities as representation of the
gray zone between A and B. The same is expected to hold in a quantum gravity
that acts as the covering space to all possible spacetimes. In certain sense
this is similar to Feynman's path integral picture where each path has
certain probability. Why is interpolation / entanglement of A and B so
important? We recall that wave-particle duality of quantum theory allows
states such as 30\% particle 70\% wave and so on. Through inherent
non-linearity general relativity provides an arena in which all sources
contribute to the resulting spacetime. The mathematical procedure developed
by Riesz \cite{13} covers the simplest two-level systems which have two
fixed end states interpolated by a single parameter. Clearly our spacetimes
at hand are not at the level of bits or qubits, but the method of classical
interpolation bears traces of reminiscences to particle state superposition.
It is not difficult to anticipate that as superposition law lies at the
heart of quantum theory interpolation of spacetimes may play a similar role
in a futuristic quantum gravity. Mathematically we admit that interpolation
is not a unique process; different parametrization leads to different gray
zone spacetimes.

In this study by resorting to the Riesz potential / interaction approach 
\cite{13} for two bodies we wish to reinterpret the analogous problems in
general relativity. Our aim can be summarized with the following examples.

A) The Coulomb potential is given by $V\left( r\right) =\frac{Q}{r},$ for
charge $Q.$ Following Riesz \cite{13} we redefine this potential by 
\begin{equation}
V\left( r,s\right) =\frac{1}{s}\left( 1-e^{-\frac{Qs}{r}}\right)
\end{equation}%
for the parameter $0\leq s<\infty .$ Obviously this potential interpolate
the Coulomb field ($s\rightarrow 0$) with the vacuum ($s\rightarrow \infty $%
). Thus (1) acts as a screening potential with the screening parameters for
the Coulomb field. (Note that for $Q<0$ we choose $-\infty <s\leq 0.$)

B) The static Schwarzschild (S) metric is given by%
\begin{equation}
ds^{2}=-f\left( r\right) dt^{2}+\frac{dr^{2}}{f\left( r\right) }+r^{2}\left(
d\theta ^{2}+\sin ^{2}\theta d\varphi ^{2}\right)
\end{equation}%
with $f\left( r\right) =1-\frac{2m}{r}.$ We modify this line element now by 
\begin{equation}
f\left( r,s\right) =1+\frac{1}{s}\left( e^{-\frac{2m}{r}s}-1\right)
\end{equation}%
where $0\leq s<\infty $ is a constant parameter that interpolates between a
S black hole and flat spacetime. Indeed, we have 
\begin{equation}
\lim_{s\rightarrow 0}f\left( r,s\right) =1-\frac{2m}{r}
\end{equation}%
and%
\begin{equation}
\lim_{s\rightarrow \infty }f\left( r,s\right) =1.
\end{equation}%
To have a black hole solution we must choose $0\leq s<1,$ whereas the
particular case $s=1$ matches with the S solution in the first order of
expansion.

The source created by the parameter $0<s<\infty $ deserves a separate study
which is out of our scope in this paper. The parameter $s$ may be
interpreted as a screening parameter for the S black hole. For $s>1,$ the
solution loses its black hole property, and for $s\gg 1$ we approach to the
flat (or vacuum) spacetime.

C) In a similar manner we consider the charged BTZ spacetime with zero
angular momentum in $2+1-$dimensions%
\begin{equation}
ds^{2}=-g\left( r\right) dt^{2}+\frac{dr^{2}}{g\left( r\right) }%
+r^{2}d\varphi ^{2}
\end{equation}%
where%
\begin{equation}
g\left( r\right) =-M+\left( \frac{r}{\ell }\right) ^{2}-\frac{Q^{2}}{2}\ln
\left( \frac{r}{r_{0}}\right) .
\end{equation}%
The parameters $M,\ell $ and $Q$ are related to mass, cosmological constant
and electric charge, respectively, while $r_{0}$ is a scaling constant (with 
$r>r_{0}$).

As in the S case (B) above, and following the Riesz potential approach we
revise the BTZ spacetime metric function accordingly as 
\begin{equation}
g\left( r,s,p\right) =-M+\frac{1}{s}\left( 1-e^{-s\left( \frac{r}{\ell }%
\right) ^{2}}\right) +\frac{Q^{2}}{2p}\left( \left( \frac{r_{0}}{r}\right)
^{p}-1\right) .
\end{equation}%
Here $0\leq s,p<\infty $ are parameters that interpolate between the vacuum (%
$s,p\rightarrow \infty $) \ by fixing the constant $M,$ and charged BTZ ($%
s,p\rightarrow 0$) spacetime. Obviously for $Q=0,$ $s$ interpolates the
vacuum with the cosmological constant spacetime. It is seen that charged BTZ
metric provides an example of two parametric interpolation in accordance
with the Riesz's prescription.

Along similar line of thought, but excluding the flat space as a limit we
consider two basic spacetimes such as S and dS and interpolate them by a
parameter as described in the next section.

\section{Schwarzschild-de Sitter spacetime}

For many reasons the Schwarzschild (S) and de-Sitter (dS) spacetimes are two
best known / cited exact solutions in general relativity. The first (S) is a
black hole solution while the second (dS) is a cosmological solution. The
intersection / coexistence of the two is known as the Schwarzschild- de
Sitter (SdS) solution. To understand non-rotating black holes in cosmology,
their thermodynamics, particle creation etc. this solution provides a basic
reference (see for instance \cite{14}). Being so important it deserves to
revisit such a spacetime from a different perspective.

Our line element is chosen now to be%
\begin{equation}
ds^{2}=-f\left( r\right) dt^{2}+\frac{dr^{2}}{f\left( r\right) }+r^{2}\left(
d\theta ^{2}+\sin ^{2}\theta d\varphi ^{2}\right)
\end{equation}%
where 
\begin{equation}
f\left( r\right) =1-2\frac{M_{0}}{r}\sin \lambda -\left( \frac{r}{\ell _{0}}%
\right) ^{2}\cos \lambda
\end{equation}%
in which $0\leq \lambda \leq \frac{\pi }{2}$ is the interpolation parameter.
Obviously by a redefinition%
\begin{equation}
M=M_{0}\sin \lambda
\end{equation}%
and%
\begin{equation}
\frac{1}{\ell ^{2}}=\frac{1}{\ell _{0}^{2}}\cos \lambda
\end{equation}%
we have the standard SdS metric satisfying the constraint condition%
\begin{equation}
\left( \frac{M}{M_{0}}\right) ^{2}+\left( \frac{\ell _{0}}{\ell }\right)
^{4}=1.
\end{equation}

\begin{figure}[tbp]
\includegraphics[width=70mm,scale=0.7]{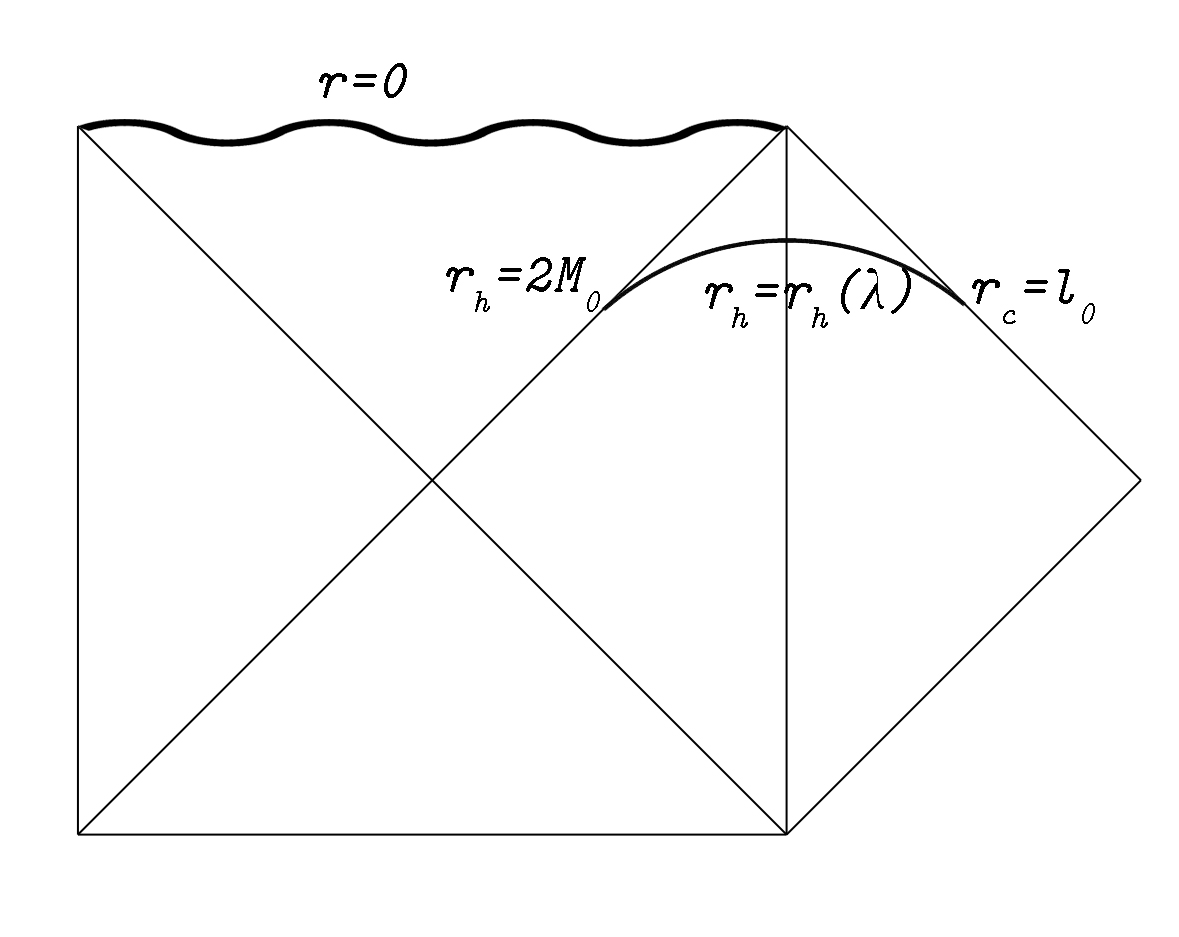} %
\captionsetup{justification=raggedright, singlelinecheck=false}
\caption{Penrose diagram for the SdS spacetime. The black hole horizon $%
r_{h}=2M_{0}$ and the cosmological horizon $r_{c}=\ell _{0}$ are
interpolated by the curve $r_{h}\left( \protect\lambda \right) .$ The
particular case $r_{h}=r_{c},$ which amounts to $\frac{M_{0}}{\ell _{0}}\sin 
\protect\lambda \protect\sqrt{\cos \protect\lambda }=\frac{1}{3\protect\sqrt{%
3}},$ corresponds to the Nariai limit. Detailed plots are given in the
following figures.}
\end{figure}

\begin{figure}[tbp]
\includegraphics[width=70mm,scale=0.7]{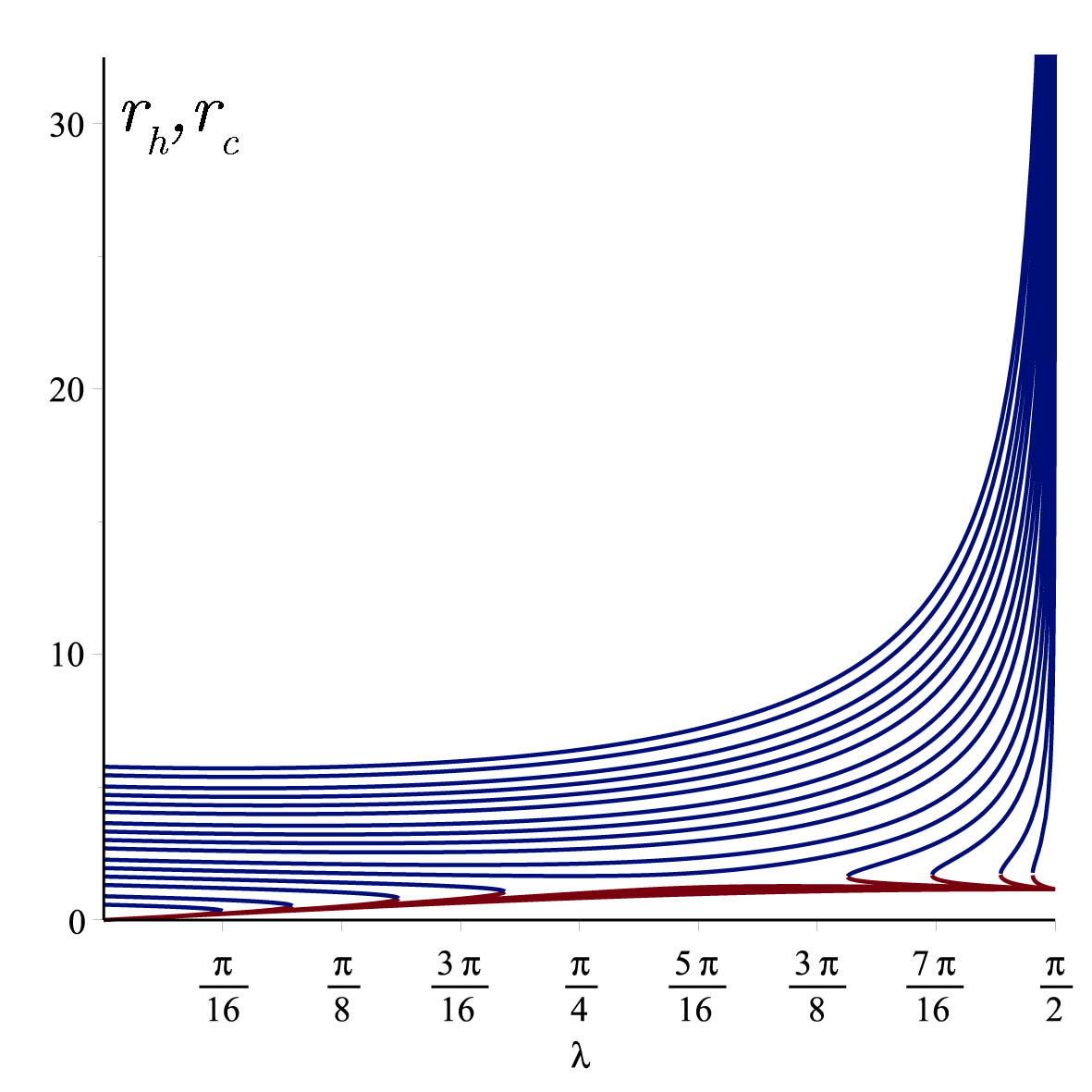} %
\captionsetup{justification=raggedright, singlelinecheck=false}
\caption{$r_{h}$ (red) and $r_{c}$ (blue) versus $\protect\lambda $ for
various $1\leq \ell _{0}\leq 10$ with $\Delta \ell _{0}=0.2.$ At $\ell
_{0}=3.224$ the horizons coincide.}
\end{figure}
\begin{figure}[tbp]
\includegraphics[width=70mm,scale=0.7]{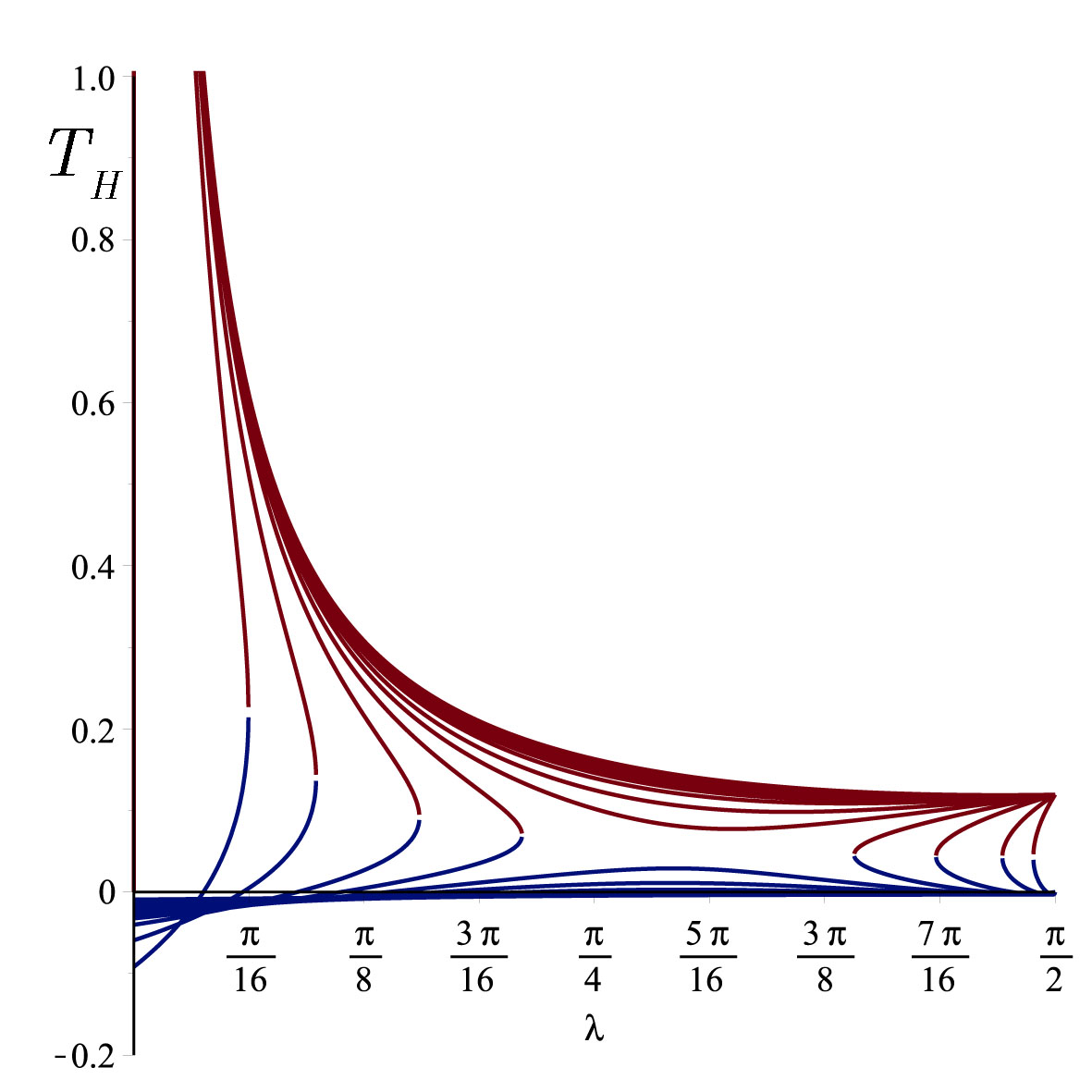} %
\captionsetup{justification=raggedright, singlelinecheck=false}
\caption{$T_{H}$ at $r_{h}$ (red) and $r_{c}$ (blue) versus $\protect\lambda 
$ for various $1\leq \ell _{0}\leq 10$ with $\Delta \ell _{0}=0.2.$ At $\ell
_{0}=3.224$ the two temperatures coincide.}
\end{figure}

In the original SdS metric the parameters $M_{0}$ and $\ell _{0}$ are
independent parameters coupled through the non-linear dynamics of general
relativity. We go one more step forward to entangle / couple them through
the parameter $\lambda .$ The interpolation parameter $\lambda $ can
appropriately be dubbed also as a mixing angle between $M_{0}$ and $\ell
_{0}.$ It can be interpreted through (11) that $\ell _{0}$ makes / dresses
the mass: the local mass $M_{0}$ is coupled with the cosmological $\ell
_{0}, $ as implied by the Mach Principle. Stated otherwise, at each point of
spacetime we have a cosmologically induced mass. Upon this arrangement the
entire dynamics of the SdS spacetime becomes dependent on the parameter $%
\lambda .$ For each choice of $0<\lambda <\frac{\pi }{2}$ we have a
spacetime with entangled $M$ and $\ell $ through (11) and (12). The Penrose
diagram for the SdS spacetime is shown in Fig. 1. Einstein's equations are
summarized as 
\begin{equation}
G_{\mu }^{\nu }=R_{\mu }^{\nu }-\frac{1}{2}R\delta _{\mu }^{\nu }=T_{\mu
}^{\nu }
\end{equation}%
where 
\begin{equation}
T_{\mu }^{\nu }=\left( \rho +p\right) u_{\mu }u^{\nu }+p\delta _{\mu }^{\nu }
\end{equation}%
so that with the choice $u^{\mu }=\delta _{0}^{\mu },$ 
\begin{equation}
\rho =\frac{3\cos \lambda }{\ell _{0}^{2}}>0
\end{equation}%
and%
\begin{equation}
p=-\frac{3\cos \lambda }{\ell _{0}^{2}}<0
\end{equation}%
and the energy condition 
\begin{equation}
\rho +p=0
\end{equation}%
holds. The fact that $\rho >0$ justifies our choice of the interpolation
parameter in the interval $0<\lambda <\frac{\pi }{2}.$ Accordingly the
scalar curvature is 
\begin{equation}
R=\frac{12\cos \lambda }{\ell _{0}^{2}}
\end{equation}%
and the Kretschmann scalar becomes%
\begin{equation}
K=24\left( \frac{\cos ^{2}\lambda }{\ell _{0}^{2}}+\frac{2M_{0}^{2}\sin
^{2}\lambda }{r^{6}}\right) .
\end{equation}%
All the physical properties of SdS spacetime are also interpolated. The
double horizons i.e., event and cosmological for instance, from $f\left(
r\right) =0,$ yields%
\begin{equation}
r_{h}=\frac{2\ell _{0}}{\sqrt{3}\cos \lambda }\cos \left( \frac{\psi +\pi }{3%
}\right)
\end{equation}%
and%
\begin{equation}
r_{c}=\frac{2\ell _{0}}{\sqrt{3}\cos \lambda }\cos \left( \frac{\psi -\pi }{3%
}\right)
\end{equation}%
in which 
\begin{equation}
\cos \psi =3\sqrt{3}\left( \frac{M_{0}}{\ell _{0}}\right) \sin \lambda \sqrt{%
\cos \lambda }.
\end{equation}%
Fig. 2 plots the graphs of $r_{h}$ and $r_{c}$ in terms of $\lambda $ and
various values of $\ell _{0}.$We recall that $r_{h}$ and $r_{c}$ are the
black hole and cosmological horizons, respectively. In other words, it can
be checked that for $\lambda \rightarrow 0$ we get $r_{c}=\ell _{0},$ and
for $\lambda \rightarrow \frac{\pi }{2}$ it yields $r_{h}=2M_{0}$, as it
should. Let us note that for real roots we impose the condition 
\begin{equation}
\left( \frac{M_{0}}{\ell _{0}}\right) ^{2}\cos \lambda \sin ^{2}\lambda \leq 
\frac{1}{27}.
\end{equation}%
For the case $\psi =0,$ the two horizons coincide (i.e., $r_{h}=r_{c}=\frac{%
\ell _{0}}{\sqrt{3}\cos \lambda }$) at the Nariai horizon \cite{15,16}.
Naturally it has to be imposed also that $\ell _{0}>2M_{0}$. The Hawking
temperatures at $r_{h}$ and $r_{c}$ are given respectively by%
\begin{equation}
T_{H}=\frac{1}{2\pi }\left\vert \frac{M_{0}\sin \lambda }{r_{h}^{2}}-\frac{%
r_{h}\cos \lambda }{\ell _{0}^{2}}\right\vert
\end{equation}%
and 
\begin{equation}
T_{H}=\frac{1}{2\pi }\left\vert \frac{M_{0}\sin \lambda }{r_{c}^{2}}-\frac{%
r_{c}\cos \lambda }{\ell _{0}^{2}}\right\vert
\end{equation}%
which are depicted in Fig. 3.

\section{Conclusion}

Schwarzschild (S) and de Sitter (dS) spacetimes are interpolated / connected
by the finite parameter $0<\lambda <\frac{\pi }{2}$, which constrains the
sources of both spacetimes. The coupling automatically rules the two
spacetimes by the common parameter $\lambda .$ Accordingly the horizons,
Hawking temperatures and other physical properties are not independent any
more, but interpolated as well. Through interpolation the effect of
cosmological constant at large is coupled and felt at small with mass and
vice versa. In certain sense the two spacetimes become ever coupled in
reminiscence with the particle entanglement encountered in quantum theory.
With the example of SdS the infinite ($0\leq s<\infty $) range interpolation
of Riesz is extended to a finite range. The example of BTZ suggests that
each physical parameter is interpolated independently, i.e., with two
parameters $s$ and $p.$ The same rule applies also to Reissner-Nordstr\"{o}m
space and vacuum. As stated above the infinite range parameter of Riesz has
been extended to a finite range, and also to the case of more than one
parameter. In brief, we propose the interpolation method of two spacetimes,
(or in case of two particles) as an alternative to the connection through a
wormhole. We add finally that the scope of the technique is not limited by
these examples.\ Given that the process of interpolation is non-unique it
can be enriched easily with further applications.

\bigskip

\end{document}